\documentclass[aps,prx,twocolumn,superscriptaddress,longbibliography,nobalancelastpage]{revtex4-2}

\pdfoutput=1

\usepackage{epstopdf}

\usepackage{graphicx}
\usepackage{amsmath,amssymb,amsfonts}
\usepackage[colorlinks,citecolor=blue,linkcolor=blue,urlcolor=blue, breaklinks=true]{hyperref}
\usepackage{color}
\usepackage[normalem]{ulem}
\usepackage{float}

\usepackage[T1]{fontenc}

\usepackage{physics} 
\usepackage[percent]{overpic} 
\usepackage{bm}

\usepackage[english]{babel}

\usepackage{xcolor}
\usepackage{xspace}
\usepackage{ifthen}

\newcommand{\lucia}[1]%
{\ifthenelse{\equal{\showcomments}{true}}%
{{\color{orange}{\small \textbf{A:} #1}}}{\xspace}}%

\newcommand{\richard}[1]%
{\ifthenelse{\equal{\showcomments}{true}}%
{{\color{purple}{\small \textbf{R:} #1}}}{\xspace}}%

\newcommand{\showcomments}{true}

\usepackage[normalem]{ulem}

\begin{document}

\title{Gravitationally-induced entanglement in 
 cold atoms}

\author{Richard Howl}
\email{richard.howl@rhul.ac.at}
\affiliation{Department of Physics, Royal Holloway, University of London, Egham, Surrey, TW20 0EX, United Kingdom}

\author{Nathan Cooper}
\affiliation{School of Physcis and Astronomy, University of Nottingham, University Park, Nottingham, NG7 2RD, UK}

\author{Lucia Hackerm\"{u}ller}
\affiliation{School of Physcis and Astronomy, University of Nottingham, University Park, Nottingham, NG7 2RD, UK}


\date{\small\today}

\begin{abstract} \noindent 
A promising route to testing quantum gravity in the laboratory is to look for gravitationally-induced entanglement (GIE) between two or more quantum matter systems. Proposals for such tests have principally used microsolid systems, with highly non-classical states, such as N00N states or highly-squeezed states. Here, we consider, for the first time, GIE between two  atomic gas interferometers as a test of quantum gravity. We propose placing the two  interferometers next to each other in parallel and looking for correlations in the number of atoms at the output ports as evidence of GIE and quantum gravity. GIE is possible without challenging macroscopic superposition states, such as N00N or Schr\"{o}dinger cat states, and instead there can be just classical-like `coherent' states of atoms. This requires the total mass of the  atom interferometers to be on the Planck mass scale, and long integration times. However, with current state-of-the-art quantum squeezing in cold atoms, we argue that the mass scale can be reduced to approachable levels and  detail  how such a mass scale can be achieved in the  near future.
\end{abstract}

\maketitle

\section{Introduction}

The last few years have witnessed an explosion of interest in testing quantum gravity with small-scale experiments \cite{huggett2022quantum}. The general idea is to, rather than bringing gravity to the traditional microscopic scales of quantum theory, such as using particle accelerators operating at  energies that appear far out of reach \cite{hossenfelder2018lost}, to instead bring quantum theory to the macroscopic realm of gravity. This approach dates back to a thought experiment of Feynman \cite{FeynmanQG}, where a massive quantum object is placed near another object that has been put in a superposition of two locations by a Stern-Gerlach experiment. If gravity obeys quantum  theory then we would expect the gravitational field or force of the superposed object to also be in a superposition, resulting in entanglement between the two quantum objects \cite{bose2017spin,marletto2017gravitationallyinduced}. However, if gravity obeys classical theory, then we would not naturally expect entanglement between the objects. The experiment is, therefore, a test of whether gravity is quantized.

Most proposals for such table-top tests of quantum gravity have followed Feynman's in that they look for gravitationally-induced entanglement (GIE) between two massive, solid objects \cite{bose2017spin,marletto2017gravitationallyinduced,krisnanda2017revealing,Qvarfort_2020,PhysRevLett.128.110401,PhysRevA.101.063804,Datta_2021,plato2022enhanced}. This is due, in part, to the theoretical simplicity of such experiments, and also because the field has seen rapid developments, such as the cooling of a microsolid to its ground state \cite{delic2020cooling}. However, it has been noted, see e.g.\ \cite{Aspelmeyer2022}, that these proposals require huge advancements over current state-of-the-art, such as in the  quintessential Bose et al., proposal \cite{bose2017spin} where, since only single atoms have been sent through Stern-Gerlach experiments, a new technological leap is required \cite{PhysRevA.107.032212} to use microsolids with masses nine to ten orders of magnitude  greater.

However, an experimental field that has been  overlooked for GIE tests is cold atomic gases. As pointed out in the pioneering work of  Marletto and Vedral \cite{marletto2017gravitationallyinduced},  the idea of GIE can be extended to many physical systems, including cold atoms. However, the example test proposed in \cite{marletto2017gravitationallyinduced} again involved a microsolid, and so far there has been no concrete experimental proposal that explicitly looks for and measures  GIE in cold atomic gases. Instead, discussions on GIE with atomic gases have been limited to a proposal \cite{carney2021using} where an atomic gas is used as the probe for inferring rather than directly measuring the entanglement with a massive solid object in a superposition of locations, which has been shown to not be a robust enough test of quantum gravity \cite{universe8020058,PhysRevResearch.4.013024,PhysRevResearch.4.013023,carney2021comment}. It has also been argued 
 \cite{Haine_2021} that particle entanglement can be generated in a single atomic interferometer due to quantum gravity if the initial state is classical. However, the concrete experimental scheme that was proposed did not explicitly measure entanglement as evidence of quantum gravity, and starts with an initial state that is highly entangled in the particle basis.

This lack of cold-atom proposals for GIE is perhaps surprising given the advancements in cold atom tests in recent years, and that this is a well-established field for quantum technology that has reached industrial applications \cite{bongs2019taking}. For example, superpositions over a distance of  $54\,\mathrm{cm}$ have been achieved in cold atomic gases \cite{kovachy2015quantum,PhysRevLett.118.183602}, as well as quantum squeezing and entanglement \cite{Chapman2012,Oberthaler2014,hosten2016measurement,Treutlein2018,Klempt2018,Oberthaler2018}; they also offer highly-repeatable experiments, and  have been found to provide remarkable test-beds of classical gravity, such as tests of the equivalence principle \cite{Zhou2015}, Newton's constant  and in gravimetry \cite{Tino_2021}. One explanation for the lack of investigations is that it was thought that the states required to test quantum gravity needed to be very  non-classical \cite{bassi2017gravitational}, such as N00N or Schr\"{o}dinger cat states, which appear harder to achieve in cold atoms than microsolids, or that the mass density required to test quantum gravity needed to be that of microsolids and thus unattainable to cold atom experiments \cite{Aspelmeyer2022}.

Here, we consider a cold-atomic gas proposal for GIE, demonstrating that cold atoms provide promising systems for such tests,  with no need for N00N states or densities of microsolids. 

\section{Proposal}

The proposed experiment consists of  two atomic interferometers in the parallel Mach-Zehnder configurations of Figure \ref{fig:Int} \cite{nguyen2020entanglement}. The gravitational force between the atoms of one interferometer and the other induces entanglement between the  interferometers, which is revealed by correlations in the number of atoms in the output ports of the interferometers. Physical implementations of the   interferometers could be atomic fountains with free-falling atomic clouds \cite{Tino_2021} or trapped interferometers, such as cold atoms in double-well potentials \cite{PhysRevLett.92.050405,schumm2005matter,PitaevskiiBook,cronin2006atom,PhysRevLett.94.090405} or optical lattices \cite{panda2022quantum}. We leave the physical implementation of the atom interferometer open, demonstrating how the parameter space is wide enough for the implementation of any of the above schemes. However, owing to their expected longer interaction times, we pay extra attention to a double-well setup, Figure \ref{fig:DW}, which, as with atom fountains, can be used with thermal or condensed atoms \cite{Dupont-Nivet_2016,PhysRevA.91.053623,Dupont-Nivet_2018}. 

As well as the gravitational interaction, the interferometers will also  interact through electromagnetism. However, those interactions can be neglected by   making sure that the interferometers are sufficiently far apart such that the gravitational interaction dominates over the  electromagnetic interactions between neutral atoms \cite{bose2017spin} or by placing a conducting membrane between the interferometers \cite{PhysRevA.102.062807,westphal2021measurement,Schmole_2016}. This is the principal reason why we consider two atomic interferometers for a test of quantum gravity rather than one, as in \cite{Haine_2021}. Additionally, there is the issue of whether, from a theory  point-of-view \cite{marletto2017gravitationallyinduced,bose2017spin,marletto2020witnessing,Galley2022nogotheoremnatureof}, local classical gravity could still create entanglement in a single interferometer through direct interactions as the atoms are brought together in the beam splitters, and the fact that, for the initial state to be non-classical, it must already be entangled in the particle basis, while, in the mode basis, entanglement is not generated by quantum gravity when the initial state is separable (see below). These issues are not present when using  two interferometers.

Although the setup in Figure \ref{fig:Int} is similar to that often seen in microsolid GIE proposals \cite{bose2017spin,marletto2017gravitationallyinduced}, in those experiments the microsolid is in a superposition of two paths and is thus in a Schr\"{o}dinger cat or N00N state, where N is the number of atoms in each object. In contrast, the atoms in Figure \ref{fig:Int} are in independent superpositions, such that there is no N00N state and instead a far more classical-like state, often referred to as a `coherent state', which are much simpler and  routinely produced in atom interferometer experiments \cite{PitaevskiiBook}. This significant difference has been the source of some confusion in the literature \cite{kovachy2015quantum,stamper2016verifying,kovachy2016response}.

\subsection{Initial state}

\begin{figure}
    \centering
    \includegraphics[width=0.3\textwidth]{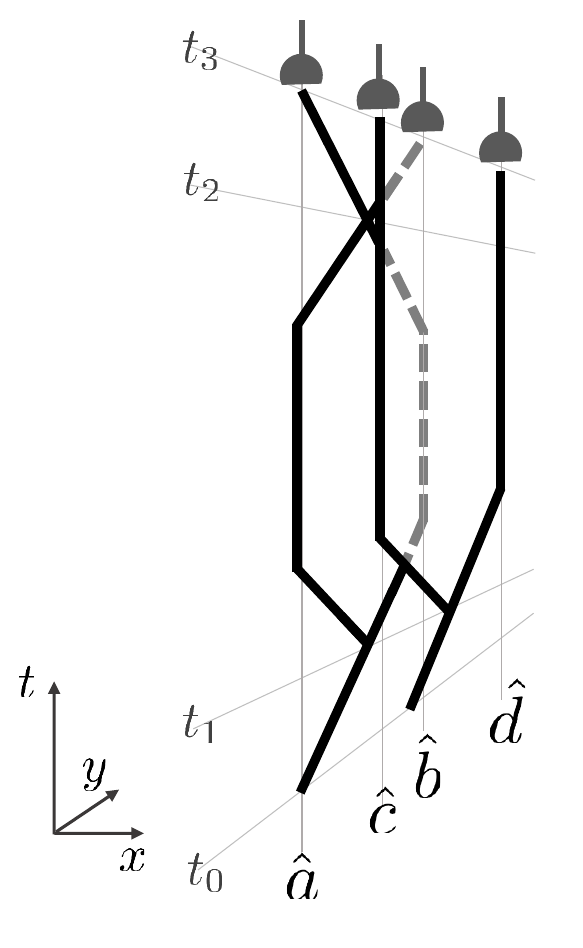}
    \caption{Two atom interferometers (one labelled by modes $\hat{a}$ and $\hat{b}$, and the other by $\hat{c}$ and $\hat{d}$) are placed adjacent to each other and in parallel \cite{nguyen2020entanglement} (gray dashed lines indicate one interferometer behind the other).  The interferometers are far enough apart, or a thin conducting sheet is placed between them, such that they 
    only interact with each other  gravitationally. At time $t_1$  the atoms of both interferometers are beam splitted into two arms, after which they interact with the atoms of the other interferometer. At time  $t_2$ the arms of the $ab$ interferometer are recombined, whereas the other interferometer arms are left open. Finally, at $t_3$ the number of atoms are measured, for example with single-atom detectors \cite{bakr2009quantum,Bakr547,sherson2010single,perrin2012hanbury,Brennecke11763,RevModPhys.85.553,PhysRevLett.109.220401,PRXQuantum.2.010325}. Correlations in the differences in the number of atoms in each output port are sought as evidence of the quantum nature of gravity (see \eqref{eq:S}).}
    \label{fig:Int}
\end{figure}

The first stage of the interferometer, $t_1$ in Figure \ref{fig:Int}, beam splits the atoms. Physical implementations of this include Bragg diffraction \cite{PhysRevLett.100.180405}, modifying the potential barrier of a double-well potential \cite{PhysRevLett.78.4675,RevModPhys.73.307,PitaevskiiBook}. We take the beam splitter to be a 50-50 beam splitter, and assume that each interferometer is in a `coherent' state with the full system   a product of coherent states $|\psi\rangle = |\phi\rangle_{ab} \otimes |\phi'\rangle_{cd}$, where \cite{PitaevskiiBook}:
 \begin{align} \label{eq:initialState}
 |\phi \rangle_{\alpha \beta} &:= \frac{1}{\sqrt{2^N}} \sum^N_{k = 0} \sqrt{\left( \begin{array}{c} N \\ k \end{array} \right)} e^{i k \phi} |N-k \rangle_{\alpha} |k \rangle_{\beta},
\end{align}
with $\alpha,\beta = \{a,b,c,d\}$  the modes of the interferometers as labelled in Figures \ref{fig:Int} and \ref{fig:DW};  $N$ the number of atoms in each interferometer, which we assume to be the same just for mathematical convenience \footnote{This assumption is only made for mathematical simplicity, the scheme will also work if the total number of atoms in each interferometer  differ.}; and $\phi$ and $\phi'$ the phases induced by the beam splitters: The state \eqref{eq:initialState} is  created by the beam splitter unitary operator $\hat{U}_{\alpha \beta}(\theta,\phi) := \exp[ \pi/4  ( e^{i \phi } \hat{\alpha} \hat{\beta}^{\dagger} - e^{-i \phi }\hat{\alpha}^{\dagger} \hat{\beta})]$ acting initially on $N$ atoms in one mode \cite{KnightBook}. 

Here we are describing the interferometers using second-quantization (the mode basis). In first-quantization (the particle basis), the coherent state is:
\begin{align}
    |\phi\rangle_{\alpha \beta} = \frac{1}{\sqrt{2^N}} \bigotimes^N_{i=1}\, \Big(|\alpha\rangle_i + e^{i \phi} |\beta\rangle_i\Big),
\end{align}
where $i$ labels the particles and $\alpha$ and $\beta$ the modes. This  emphasizes that the coherent state is just a product of independent atoms each in a superposition.

\subsection{Evolved state}

\begin{figure}
    \centering
    \includegraphics[scale=0.4]{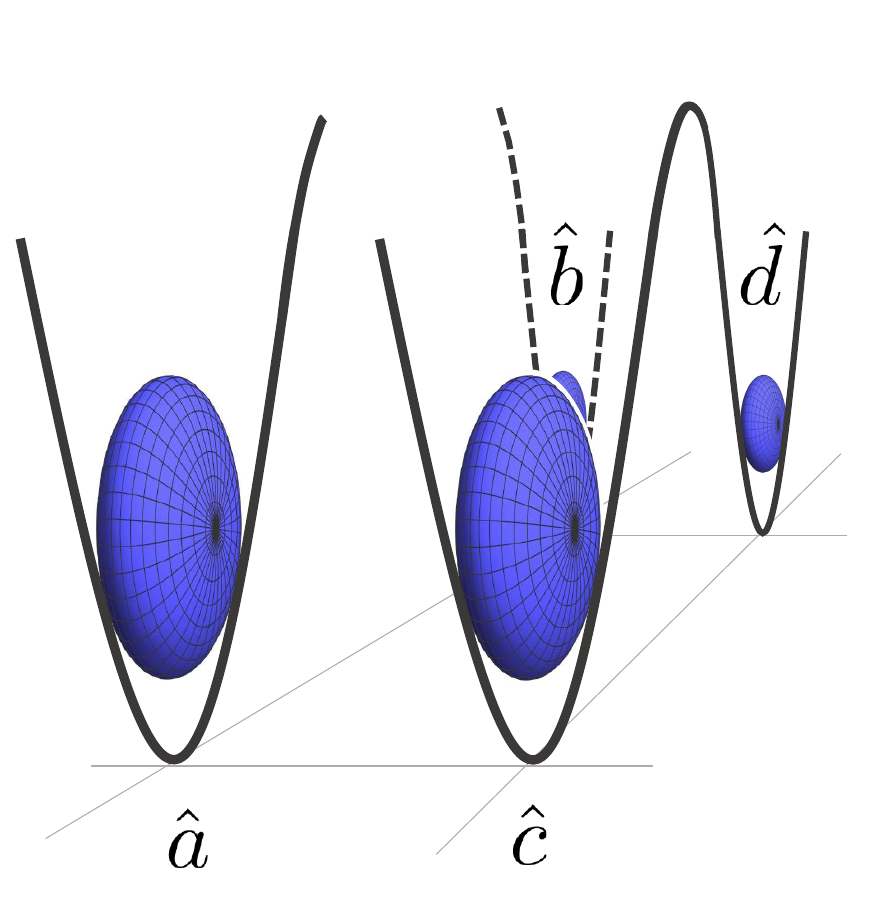}
    \caption{An example of a physical realization of the interferometry scheme of Figure \ref{fig:Int}. Two adjacent double-well harmonic potentials store atomic clouds with  oblate spheroid (pancake) geometries. The atoms can be beam-splitted by lowering the double-well potential barrier, which creates a Josephson junction. One double well is labelled by modes $\hat{a}$ and $\hat{b}$,  the other by $\hat{c}$ and $\hat{d}$. The distance between the two interferometers is closer than the distance between each arm of the interferometer to enhance the correlations between the interferometers due to their gravitational interaction.}
    \label{fig:DW}
\end{figure}

The atoms are now left to interact via gravity. In the non-relativistic (Newtonian) limit of quantum gravity, this interaction is described by a Hamiltonian of the form \cite{PRXQuantum.2.010325}:
\begin{align} 
\hat{H} &= -\frac{1}{2} G \int d^3 \bm{x} d^3 \bm{x}' \sum_{\alpha,\beta = \{a,b,c,d\}}   \frac{:\hat{\rho}_{\alpha}(\bm{x}) \hat{\rho}_{\beta}(\bm{x}'):}{|\bm{x} - \bm{x}'|}\\ \label{eq:HQG}
&=:  \sum_{\alpha,\beta = \{a,b,c,d\}}  \lambda'_{\alpha \beta } \hat{\alpha}^{\dagger}  \hat{\beta}^{\dagger} \hat{\alpha} \hat{\beta}, 
\end{align}
 where $:\,:$ represents normal ordering, $\hat{\rho}_{\alpha}(\bm{x}) := m\hat{\Psi}^{\dagger}_{\alpha}(\bm{x}) \hat{\Psi}_{\alpha}(\bm{x})$ is the mass-density operator for the atoms, $\hat{\Psi}_{\alpha}(\bm{x})$ is the non-relativistic atomic field, and we have approximated the wavefunctions of each mode to be non-overlapping and used a single-mode approximation \cite{PitaevskiiBook}: $\hat{\Psi}_{\alpha}(\bm{x}) \approx \psi_{\alpha} (\bm{x}) \hat{\alpha}$, with $\hat{\alpha}$ the annihilation operator for the corresponding mode (satisfying $[\hat{\alpha},\hat{\alpha}^{\dagger}] = 1$)  and $\psi_{\alpha}(\bm{x})$ the `wavefunction' of the mode.  For $\lambda'_{\alpha \beta}$, we have:
\begin{align} \label{eq:lambdap}
\lambda'_{\alpha \beta} &:= - \frac{1}{2} G m^2 \int d^3 \bm{x} d^3 \bm{x}' \frac{|\psi_{\alpha}(\bm{x}')|^2 |\psi_{\beta}(\bm{x}')|^2 }{|\bm{x}- \bm{x}'|},
\end{align}
where $m$ is the mass of the atoms and $G$ Newton's constant. To obtain a concrete expression for $\lambda'_{\alpha \beta}$, we assume a trapped double-well system with a standard harmonic trapping potential for each well. Then, to 
maximize the strength of the gravitational interaction between the interferometers, we assume harmonic potentials that force the atomic clouds to have oblate spheroid wavefunctions (see Figure \ref{fig:DW}). This is found to provide a small enhancement of  $\approx 3/2$  over using spheres of the same volume - see Appendix \ref{app:oblate}. In a non-trapped scheme, $\lambda'_{\alpha \beta}$ would need to be calculated for the relevant wavefunctions $\psi_{\alpha} (\bm{x})$ of the particular setup in that case.

Given the configuration of Figure \ref{fig:Int}, we ignore the interaction between the two modes within the interferometer ($a$ with $b$ and $c$ with $d$) and the diagonal modes between the two interferometers  ($a$ with $d$ and $b$ with $c$) \cite{nguyen2020entanglement}, such that the only sizeable gravitational interactions are the self-interactions of each mode, and between the adjacent modes $a$ and $c$, and $b$ and $d$ of the two interferometers. We also assume that all modes are identical, such that $ \lambda'_{aa} = \lambda'_{bb} = \lambda'_{cc} = \lambda'_{dd} =: \lambda'_s$ and $\lambda'_{ac}=\lambda'_{bd}:=\lambda'$. The state of the system after the atoms have interacted gravitationally for time $t$ is then  \footnote{We assume that, since the atoms are trapped, we can neglect the movement of the atoms due to their gravitational interaction.}:
 \begin{align}\nonumber
   |\psi_t \rangle &= \frac{1}{2^N}  \sum^N_{k,k' = 0} \sqrt{\left( \begin{array}{c} N \\ k \end{array} \right)\left( \begin{array}{c} N \\ k' \end{array} \right)} e^{i (k \Phi +k' \Phi')   }  e^{i \lambda_s(k^2 + k^{'2})}\\ \label{eq:StateGravityInt}
    &\hspace{1.5cm} \times e^{2 i \lambda k k'} 
    |N-k \rangle_a |k \rangle_b  |N-k' \rangle_c |k' \rangle_d,
\end{align}
where $\Phi := \phi - N(\lambda_s + \lambda)$ and $\Phi' := \phi' - N(\lambda_s + \lambda)$, with $\lambda_s := 2\lambda'_s t / \hbar$ and $\lambda :=2 \lambda' t / \hbar$.   
In \eqref{eq:StateGravityInt}, we have ignored the electromagnetic interactions between the atoms within each interferometer, and will introduce these later. For convenience, we have also ignored the free evolution, which will just contribute to the phases $\Phi$ and $\Phi'$.

\subsection{Entanglement and measurement}

From \eqref{eq:StateGravityInt}, the gravitational interaction will in general entangle the two interferometers. In the neglection of decoherence, We can characterize this using the concurrence, an entanglement monotone that has recently been extended to multi-particle and continuous-variable states as \cite{PhysRevA.64.042315,bhaskara2017generalized,PhysRevA.105.052441}:
\begin{align} \label{eq:C}
    C(\rho) = \sqrt{2 (1 - \mu_R)},
\end{align}
where $\mu_R$ is the purity of the reduced density matrix across the appropriate bi-partition of the pure state. In Appendix \ref{app:purity}, using \eqref{eq:StateGravityInt}, we find the purity of the state of one of the interferometers to be:
\begin{align}
    \mu 
    &= \frac{1}{2^{2N}} \sum^{2N}_k \left( \begin{array}{c} 2N \\ k \end{array} \right) \cos^{2N}[ \lambda (N-k)]. 
\end{align}
In the assumption that $\lambda N \ll 1$, this approximates $\mu \approx 1 - \frac{1}{2} \lambda^2 N^2$, and thus the concurrence is $C(\rho)\approx \lambda N$.

In practice, however, it would be hard to measure the concurrence, and \eqref{eq:C} is also only valid for pure states \footnote{An extension of this measure to mixed states has recently been considered \cite{PhysRevA.105.052441}}. Instead, to infer the entanglement in the experiment we consider closing the $ab$ interferometer by acting with another 50-50 beam splitter, and then measuring the number of atoms in each mode of the two interferometers, looking for the correlations:
\begin{align} \label{eq:S}
S:= \langle \hat{J}^{ab}_z \hat{J}^{cd}_z\rangle - \langle \hat{J}^{ab}_z\rangle \langle \hat{J}^{cd}_z\rangle,
\end{align}
where $\hat{J}^{\alpha \beta}_z := \frac{1}{2} (\hat{N}_{\alpha} - \hat{N}_{\beta})$ and $\hat{N}_{\alpha} := \hat{\alpha}^{\dagger} \hat{\alpha}$ is the number operator of the respective mode. In Appendix \ref{sec:closed}, we also consider the above correlations when both interferometers are closed. For the desired precision, looking for these atom-number correlations would likely require single-atom detectors (see Figure \ref{fig:Int}), which have been realized for cold atoms \cite{bakr2009quantum,Bakr547,sherson2010single,perrin2012hanbury,Brennecke11763,RevModPhys.85.553,PhysRevLett.109.220401,PRXQuantum.2.010325}.

With the assumption that the full state remains pure, it is clear that an $S\neq 0$ implies  entanglement since a product state results in $S=0$. However, in Appendix \ref{app:EntMeasure} we show that $S $ is a more general measure of entanglement than just for the assumption of products of coherent states and  also applies, for example, when there is a symmetrical decoherence mechanism acting on the interferometers, which is natural given the symmetry of the experimental setup (Figure \ref{fig:Int}). 
A fully robust measure of entanglement, such as potentially the cold atom analogue of the microsolid spin-entanglement witness \cite{PhysRevA.102.022428}: $W = \hat{N}_{ab}/2 \otimes \hat{N}_{cd}/2 - \hat{J}^{ab}_x \otimes \hat{J}^{cd}_x - \hat{J}^{ab}_z \otimes \hat{J}^{cd}_y - \hat{J}^{ab}_y \otimes \hat{J}^{cd}_z$, with $\hat{N}_{\alpha \beta}:= \hat{N}_{\alpha} + \hat{N}_{\beta}$ and $\hat{J}^{ab}_x$ and $\hat{J}^{ab}_y$ defined below, will be considered in future work.  

Given that $\hat{J}_{\varphi}^{\alpha \beta} = \hat{U}^{\dagger}(-\pi/4,\varphi)  \hat{J}^{\alpha \beta}_z \hat{U}(-\pi/4,\varphi)$, where $\hat{J}_{\varphi}^{\alpha \beta} = (\hat{a} \hat{b}^{\dagger} e^{i \varphi} + \hat{a}^{\dagger} \hat{b} e^{-i \varphi})/2$, the correlation \eqref{eq:S} is equivalent to 
\begin{align}
S &= \langle \psi_t |\hat{J}_{\varphi}^{ab} \hat{J}^{cd}_z|\psi_t\rangle - \langle \psi_t| \hat{J}_{\varphi}^{ab}|\psi_t \rangle \langle \psi_t| \hat{J}^{cd}_z |\psi_t \rangle,\\ \label{eq:fullS}
&= \frac{1}{4} N^2 \sin(\phi - \varphi) \cos^{N-1} \lambda_s \, \cos^{N-1} \lambda \, \sin \lambda.
\end{align}
Here, $\hat{J}_{\varphi}^{\alpha \beta}$ is the generalized `spin' operator, with $\hat{J}_{x}^{\alpha \beta}$ and $\hat{J}_{y}^{\alpha \beta}$ defined for $\varphi = 0$ and $\pi/2$ respectively. These `spin' operators obey the $\mathfrak{su}(2)$  algebra  $[\hat{J}_{p}^{\alpha \beta},\hat{J}_{q}^{\alpha \beta}] = i \epsilon_{p q r} \hat{J}_{r}^{\alpha \beta}$, where $p,q,r \in \{x,y,z\}$ such that the scheme in Figure \ref{fig:Int} is closely related to the Bose et al., proposal \cite{bose2017spin} for microsolids. Assuming $\lambda_s \approx \lambda$ and $\lambda^2 N \ll 1$, then $S \approx \lambda N^2 \sin(\phi - \varphi)/4 $. 

\subsection{Signal-to-noise}

A better suited figure of merit for the experiment than just the signal is the signal-to-noise (SNR) ratio: $\mathrm{SNR} = |S| / \sqrt{\mathrm{Var}(S)}$ where $\mathrm{Var}(S)$ is the variance associated with $S$. Since $S$ is a covariance, $\mathrm{Var}(S)$ is \cite{cook1951bi}:
    \begin{align} \label{eq:Var}
    \mathrm{Var}(\kappa_{11}) &= \frac{1}{\mathbb{M}} \kappa_{22} + \frac{1}{\mathbb{M} - 1} \kappa_{20} \kappa_{02}+ \frac{1}{\mathbb{M} - 1} S^2\\ \label{eq:Vark11}
    &\approx \frac{1}{\mathbb{M}} \Big(\kappa_{22} + \kappa_{20} \kappa_{02} +  S^2\Big), 
\end{align}
where $\mathbb{M} \gg 1$ is the number of  repetitions of the experiment, $\kappa_{20}$ is the variance of $\hat{J}_{\varphi}^{ab}$,  $\kappa_{02}$ is the variance of $\hat{J}_{z}^{cd}$,  and $\kappa_{22}$ is the bi-variate cumulant defined in \eqref{eq:k22}. 

The full expression for the SNR is non-trivial and given in \eqref{eq:fullSNR}. However, in the assumption that $\lambda_s \approx \lambda$ and $\lambda N \ll 1$, we find the following simple expression for the optimum SNR under this approximation: $\mathrm{SNR} \approx  \sqrt{\mathbb{M}} \lambda N$, which is achieved when $\varphi = \pi/2$ so that $\hat{J}_{\varphi}^{ab} \equiv \hat{J}^{ab}_y$. Note that this $\lambda N$ scaling matches that of the concurrence entanglement monotone considered above when $\lambda N \ll 1$. This scaling is proportional to $G m^2 N t / \hbar d$, where $d$ is the distance between the centre of the interferometers, which contrasts, for example, with the $G m^2 N^2 t / \hbar d$ scaling found in Bose et al., for the size of the entangling-inducing phase. This poorer scaling in $N$ is to be expected as here the state of the atoms is far more `classical-like' than the effective N00N state of \cite{bose2017spin}. However, the $\lambda N$ scaling is superior to the scaling expected for $N$ pairs of atoms that are each independently entangled, which would be $\lambda \sqrt{N}$. The advancement comes from the fact that each $N$ atom `sees' $N$ other atoms in superposition rather than just one. Although a factor of $N$ is lost compared to a N00N state scheme, a coherent state of $N$ atoms is significantly easier to create and maintain than a N00N state of $N$ atoms, since decoherence will be far less severe. 

\subsection{Required experimental parameters}

For an observable effect, an SNR at least of order one is needed. An example of the numbers required to achieve this is $N$ of order $10^{16}$ atoms of a heavy element such as erbium or cesium; a distance $d$ of  $\mathrm{cm}$'s; an equatorial and polar radii also of $\mathrm{cm}$'s (giving a number density of order $10^{12}\,\mathrm{cm^{-3}}$); $t$ of order $10^4\,\mathrm{s}$, $\mathbb{M} \approx 10^{3}$ repetitions; and assuming that there are of order $10$ such interferometers operating independently. Comparing these numbers to the state-of-the-art in cold atom interferometer experiments, we have $t \approx 10^2\,\mathrm{s}$  for an atom interferometer with optical latices \cite{panda2022quantum},   $t \approx 10^4\,\mathrm{s}$ predicted for rubidium atoms in optical-tweezer arrays \cite{PhysRevApplied.16.034013}, and three-body damping  estimates $\approx 10^3 - 10^4\,\mathrm{s}$ - see Appendix \ref{app:3Body}. Furthermore, atoms have already been superposed over $54\,\mathrm{cm}$ \cite{kovachy2015quantum,PhysRevLett.118.183602}. However, $10^{16}$ atoms have not been achieved, with just over $10^{11}$ being produced so far in MOTs \cite{PhysRevA.90.063404} and $10^8 - 10^9$ in Bose-Einstein condensates (BECs) \cite{PhysRevLett.81.3811,doi:10.1063/1.2424439}. 

To reduce the number of atoms required or reduce the challenge of the other experimental parameters, the initial state could  be quantum squeezed. For example, a squeezing of $20\,\mathrm{dB}$ in both interferometers would reduce the required atom number to $10^{14}$ (and number density $10^{11}\,\mathrm{cm^{-3}}$) \footnote{A simple way to see that  quantum squeezing will generate the above improvements, is to first consider the perturbative regime $\lambda N \ll 1$. In this case, the only part of the noise $\mathrm{Var(S)}$ that contributes to the SNR at lowest order  is the $\lambda = 0$ part, which is  just $\kappa_{20} \kappa_{02}$ at $\lambda = 0$ in \eqref{eq:Vark11}. By definition, a squeezing of $20\,\mathrm{dB}$ in each interferometer in the desired spin direction means lowering the variances  $\kappa_{20}$ and $ \kappa_{02}$ by $100$ each, thus increasing the SNR by a factor of 100.}, with $20.4\,\mathrm{dB}$ of squeezing already achieved in cold-atom experiments \cite{hosten2016measurement}. Going two orders of magnitude beyond the state-of-the-art to  $40\,\mathrm{dB}$ of squeezing  (above $33\,\mathrm{dB}$ has been suggested for squeezing of atoms via atom-light interactions \cite{PhysRevA.103.023318}), then the numbers could be further lowered to, for example, $10^{12}$ atoms (with number density $10^{12}\,\mathrm{cm^{-3}}$), a distance $d$ of $\mathrm{mm}$'s and interrogation times of order  $10^3\,\mathrm{s}$;  or $10^{13}$ atoms with  times of order $10^2\,\mathrm{s}$.

\begin{figure}
    \centering
    \includegraphics[width=0.45\textwidth]{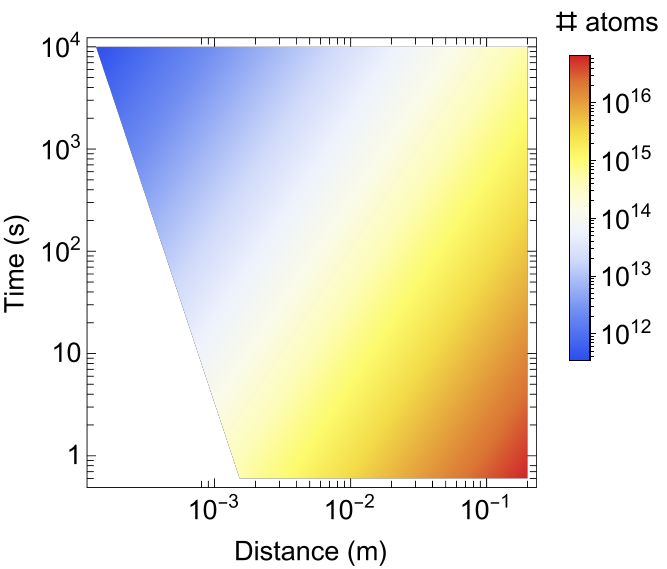}
    \caption{Parameters required to achieve an SNR of order 1 for the interferometry scheme of Figure \ref{fig:Int}.  Distance is the separation $d$ between the interferometers, Time is the interrogation time $t$, and $\#$ atoms is the total number of atoms $N$.  The atoms are erbium, squeezed at $35\,\mathrm{dB}$, $5$ independent setups are assumed, the number of repetitions are such that at $t=10^4\,\mathrm{s}$ we have $\mathbb{M}\approx10^3$, and only the parameters for which the number density $n < 10^{16}\,\mathrm{cm^{-3}}$ are shown, allowing for densities varying from MOTs to BECs (the other parameters can still be used if the  interferometry scheme is slightly modified, as discussed in the main text).}
    \label{fig:Sq}
\end{figure}

The above  parameters are just an example of what can be used to get an SNR of order 1, but the parameter space available to  an atom interferometry scheme is very large, due in part to the high tunability of such setups. The main experimental parameters involved are the interrogation time; the number of repetitions  and independent   interferometry setups; the number of atoms;  the initial squeezing; and the distance between the interferometers, which is related to the superposition size and the geometry of the clouds and interferometry setup. Figure \ref{fig:Sq}   illustrates the large parameter space available to an atom interferometry scheme, considering the distances and interrogation times available given an initial squeezing of $35\,\mathrm{dB}$ and at least $10^3$ repetitions. Only parameters for which the number density is less than $10^{16}\,\mathrm{cm^{-3}}$ are shown, and the  density varies between those of Bose-Einstein condensation to non-condensed atoms, with condensation not a necessary requirement for atom interferometry  \cite{PhysRevA.71.043615,Tino_2021}. However, even the parameters that would lead to too high densities in the scheme of Figure \ref{fig:Int} can be still achieved with small modifications to the scheme, such as using highly oblate spheroidal clouds with the interferometers in the same plane (similar to \cite{bose2017spin}) rather than parallel planes. This would reduce the density for the parameters at the expense of a loss in the gravitational interaction energy of the interferometers, but this could be preferable to a particular experimental realization, and further illustrates the high  tunability of an atom interferometry scheme and that not one particular physical realization is picked out. For example, larger superposition distances but smaller interrogation times are likely to be preferable to atom fountains, whereas the opposite landscape would be better suited to trapped atomic chip systems.

Although there is great flexibility in the numbers required,  due to the high tunability of cold atom experiments,  to see GIE, all  numbers are  challenging. This is to be expected for an experiment testing the quantum nature of gravity, and should be compared to current proposals on detecting GIE. For example, in the original, revolutionary Bose et. al., proposal \cite{bose2017spin}, microsolids of mass $10^{-15}\,\mathrm{kg} - 10^{-14}\,\mathrm{kg}$ are placed into a N00N state of superposition size $0.25\,\mathrm{mm}$ and held for $2\,\mathrm{s}$ after which entanglement is sought between their internal spin states. In comparison, the current state-of-the-art on N00N states is with molecules that are $10^{-23}\,\mathrm{kg}$ in mass, have superposition size $300\,\mathrm{nm}$ and were held for $10\,\mathrm{ms}$ \cite{fein2019quantum}, illustrating that the  \cite{bose2017spin} numbers are very challenging. Furthermore, the state-of-the-art N00N experiment with molecules uses diffraction gratings, whereas the technology proposed in \cite{bose2017spin} is with Stern-Gerlach devices, for which only  atoms and small molecules have been used \cite{doi:10.1126/sciadv.abg2879,Knickelbein2004} increasing the gap between what is required and the state-of-the-art; however a compelling case is made for such a technological leap \cite{bose2017spin,doi:10.1126/sciadv.abg2879}. An alternative is to use microscolids in traps and look for entanglement between the external position and momentum variables \cite{krisnanda2020observable,Qvarfort_2020}. This requires either highly-squeezed states, for which the technological advancements have been shown \cite{Aspelmeyer2022} to be comparable to the original   Bose et. al., proposal, or masses at least an order of magnitude above the Planck mass scale of $10^{-8}\,\mathrm{kg}$ \cite{krisnanda2020observable}.   Other schemes include looking for entanglement between massive mirrors in optical interferometers \cite{PhysRevA.101.063804,Datta_2021,plato2022enhanced}, which also tend to require masses around the Planck mass scale, dependent on the oscillator density and temperature. In comparison, the example numbers provided above for GIE in atom interferometers use of order $10^{16}$ unsqueezed erbium or cesium atoms, which is just under the Planck mass scale, but state-of-the-art $\mathrm{20}\, \mathrm{dB}$ squeezing already achieved in experiments can lower this by up to two orders of magnitude, with further squeezing presenting further improvements.

\subsection{Achieving the required atom numbers} \label{sec:MOT}

We now discuss a possible mechanism for getting to the required atom numbers, which, in all the above considered examples of the parameters needed, notably must be increased over the state-of-the-art. However, the density can be either that of thermal atoms in a  Magnetic Optical Trap (MOT) or a BEC. Since it is harder to achieve larger atom numbers in BECs, here we highlight a possible implementation with a MOT that is used to source an atom fountain or double-well interferometer, both of which can be used with non-condensed atoms at the expense of less coherence \cite{kovachy2015quantum,stamper2016verifying,kovachy2016response,PhysRevA.71.043615}.  The largest number of atoms in a MOT to date was over $10^{11}$  \cite{PhysRevA.90.063404}, however, there has been no technological drive to increase the number of atoms. We propose using a 2D MOT to load a 3D MOT. The important innovation is then the implementation of a train of slightly offset cooling frequencies, an expansion of a two-colour MOT \cite{Cao_2012,Sinclair94}. A two-colour MOT enables a larger range of capture velocities and therefore leads to an increase in the trappable atom number.  This can also be combined with other new methods and  more conventional methods, such as using larger cooling beams and dark spot MOTs. Simulations confirm that the addition of a larger number of frequencies with such methods is expected to facilitate MOTs of over $10^{14}$ atoms  \cite{FutureMOTWork}, bringing the required parameter space for testing quantum gravity within reach. 

\subsection{Experimental noise}

Above, we have considered the signal and the inherent quantum noise  of the atoms. In addition to this, there will also be experimental sources of noise. The most prominent sources of noise are expected to be the electromagnetic interactions of the atoms, collisions between the atoms and foreign atoms due to an imperfect vacuum, fluctuations in any trapping potentials, and any imperfections of the beam splitters and measuring devices.   

First, we consider the  electromagnetic interactions.  At low temperatures, the electromagnetic interactions between the atoms will be predominately two-body interactions \cite{PitaevskiiBook}: 
\begin{align}\nonumber
    \hat{H} &= \frac{1}{2} \sum_{\alpha,\beta = \{a,b,c,d\}}   \int d^3 \bm{x} d^3 \bm{x}'   V_{\alpha \beta} (\bm{x} - \bm{x'}) \\&\hspace{2.5cm}\times \hat{\Psi}^{\dagger}_{\alpha} (\bm{x}') \hat{\Psi}^{\dagger}_{\beta} (\bm{x})  \hat{\Psi}_{\alpha} (\bm{x}') \hat{\Psi}_{\beta} (\bm{x}) \\ \label{eq:HEM}
    &=:  \sum_{\alpha,\beta = \{a,b,c,d\}}  \kappa'_{\alpha \beta} \hat{\alpha}^{\dagger}  \hat{\beta}^{\dagger} \hat{\alpha} \hat{\beta},
\end{align}
where 
\begin{align}\label{eq:kappa}
\kappa'_{\alpha \beta} &:=  \frac{1}{2} \int d^3 \bm{x}  V_{\alpha \beta} (\bm{x} - \bm{x'}) |\psi_{\alpha}(\bm{x}')|^2 |\psi_{\beta}(\bm{x})|^2, 
\end{align}
and  we have assumed a one-mode approximation as we did when considering the gravitational interactions. The Hamiltonian \eqref{eq:HEM} has the same form as that of the gravitational interactions \eqref{eq:HQG}. However, we take $\kappa'_{ac} = \kappa'_{bd} \approx 0$ assuming that these can be neglected given the distance between the interferometers or that there is a  conducting membrane between them, but we keep  $\kappa'_{ab}\neq 0$ and $\kappa'_{bc}\neq 0$ in general, defining $\kappa'_d := \kappa'_{ab} = \kappa'_{cd}$. Ignoring the effect of the interactions before the beam splitter, the state of the system at time $t$ is then the same as \eqref{eq:StateGravityInt} except we now add to $\lambda_s$ the term $\kappa_s - \kappa_d$, where $\kappa_s  :=2 \kappa'_{\alpha \alpha} t / \hbar$ and $\kappa_d  := 2 \kappa'_d t / \hbar$, with $\alpha\,\beta = \{a,b,c,d\}$. Therefore, the signal and SNR are the same as \eqref{eq:fullS} and \eqref{eq:fullSNR} respectively except that we replace $\lambda_s$ with $\lambda_s+\kappa_s - \kappa_d$.

Although the electromagnetic interactions cannot create entanglement between the interferometers given the above assumptions, they can still lead to a loss of coherence \cite{PhysRevA.55.4330} and affect the signal and SNR for the gravitational  correlations. In principle, this effect can be turned off or minimized using the Feshbach resonances of the atoms \cite{PRXQuantum.2.010325,RevModPhys.82.1225,PitaevskiiBook}: Due to the Feshbach resonances, a  magnetic field of a certain strength can be applied to the gas such that the two-body electromagnetic interactions are switched off or significantly reduced \footnote{At low temperatures, the dominant two-body interactions will be s-wave interactions and magnetic dipole-dipole interactions (MDDIs). Using Feshbach resonances, the s-wave interactions can be tuned to cancel with the MDDIs \cite{PhysRevLett.101.190405,PRXQuantum.2.010325}.}, a remarkable tool not available to microsolids. Although the magnetic field strengths required to achieve this are not challenging, in order to maintain a constant zero interaction, the magnetic field must have sufficiently low noise. 


However, the electromagnetic interactions can in fact enhance the GIE effect in certain cases. For example, in the measurement scheme in Appendix \ref{sec:closed}, where both arms of the interferometer are closed, as we increase the electromagnetic interactions between the atoms (assuming the example experimental parameters  above), the signal and SNR start increasing once the strength surpasses the strength of the gravitational interactions.  This increase is maintained until the electromagnetic interactions are around three orders greater than the gravitational ones, after which, although the signal carries on increasing, the SNR starts reducing. In contrast, for the scheme in Figure \ref{fig:Int} where only one arm is closed, the signal and the SNR slightly decrease once the electromagnetic interactions  become greater than the  gravitational ones, and significantly decrease once the electromagnetic interactions are around four orders of magnitude greater than the gravitational ones. Ignoring the magnetic dipole-dipole interactions, this would mean reducing the effective s-wave scattering by around five orders of magnitude below the Bohr radius (see Appendix \ref{app:Scattering}). However, the different dependence on the strength of the electromagnetic interactions suggests that another measurement  scheme could be developed where the effect of the electromagnetic interactions is minimised further or extinguished. For example, as shown in Appendix \ref{app:purity}, the purity of each interferometer and thus the concurrence entanglement monotone considered above
 is only dependent on $\lambda$ and not on the electromagnetic interactions, although this, and its mixed state extensions \cite{PhysRevA.105.052441}, would be more challenging  to measure. Alternatively, the dependence on the electromagnetic interactions could be extracted given that the dependence of the SNR on $\gamma$ exactly matches that of measuring $\langle \hat{J}_{z}^{ab}\rangle$ in a single, isolated interferometer - see Appendix \ref{app:independentInt}.


Above, we have assumed a two-body approximation for the electromagnetic interactions. However, three-body interactions, although suppressed, will still be present and are often the limiting factor in the lifetime of trapped atomic clouds, since the collisions result in atoms being ejected out of the traps. In Appendix \ref{app:3Body}, we estimate that  three-body interactions will not be a limiting factor in measuring entanglement between the interferometers.

In addition to sufficiently reducing or distinguishing any noise due to the electromagnetic interactions of the atoms, we would also need to make sure that all other sources of noise, such as foreign atom collisions can be kept appropriately low. Such collisions will cause decoherence and are thought to be the limiting factor in GIE schemes such as Bose et.\ al., \cite{bose2017spin}, since, given the initial N00N state of atoms, a single loss of atom would be enough to completely decohere the state. However, an advantage of our scheme is that the initial state is classical-like in that each atom is, approximately, independently in a superposition. Such a coherent state, \eqref{eq:initialState}, is thus robust against single atom losses.

As well as foreign atom collisions, other sources of noise will include fluctuations of the trapping potentials and the effect of any conducting membrane used to shield the electromagnetic interactions (the membrane must be at low enough temperatures to not heat the atoms, and it may  also attract the atoms). The temperature of the gas will also cause decoherence and affect how it is split in the 50-50 beam splitter: in \eqref{eq:initialState} we assumed that the gas starts off in a single-mode Fock state, whereas in reality there will be many modes and, unless there is condensation, all the modes will be in a thermal state \cite{kovachy2015quantum}. In addition to this, it is also possible that there could be decoherence due to not closing the $cd$ interferometer - away from the Newtonian regime of quantum gravity, it is possible that the gravitational field remains entangled with matter, which would also be the case in certain other proposed tests of quantum gravity \cite{krisnanda2020observable,Qvarfort_2020}. This will be considered in future work, however, the measurement scheme considered in Appendix \ref{sec:closed} would not suffer from any such decoherence.

Another option available to cold atom systems is to operate the experiment in space, dramatically reducing noise and the challenge of achieving high interrogation times together with high atom numbers.   Cold atomic gas experiments, including atomic interferometers, have already operated in space \cite{becker2018space,aveline2020observation}, and  there are many plans to carry out further  gravitational tests with cold atoms in space \cite{alonso2022cold,cacciapuoti2020testing,Tino_2021}.


\section{Summary}

In summary, we have described a cold atoms experiment that tests quantum gravity through GIE. It is simple in that it does not require  highly-quantum states, such as generating N00N states or highly-squeezed states. However, as with any experiment to test the quantum nature of gravity, it requires advancements over the state-of-the-art - in this case, particularly the number of atoms and the interrogation time. The number of atoms required without quantum squeezing the state, means that the total mass is close to the Planck mass scale, a scale that can be used as a rough figure of merit for the ability of an experiment to test quantum gravity  \cite{penrose1990emperor,christodoulou2019possibility,Howl_2019}.  However, by slightly quantum squeezing the state, for example to current state-of-the-art levels, the mass and thus the number  of atoms required can be reduced by orders of magnitude  to approachable levels, as  detailed in Section \ref{sec:MOT}. 

The advancements required here over the state-of-the-art do not need technological leaps of faith, as with generating highly non-classical states, such as N00N states. Extending the scheme in a similar vein to novel extensions to the Bose et al., proposal \cite{bose2017spin,PhysRevA.104.052416,PhysRevA.105.032411,PhysRevD.107.064054} might also provide small improvements, such as interacting several atom interferometers and perhaps looking for multi-partite correlations rather than bi-partite correlations and entanglement. 
A key feature of cold atoms gases is that it is easy to manipulate them into different schemes and setups. Here, we have concentrated on the scheme in Figure \ref{fig:Int} where two interferometers are parallel to each other and the gasses are of oblate spheroidal shape. However, this can be tailored to the requirements of the experiment. For example, it could be necessary to have a smaller superposition size, in which case extreme oblate spheroids might be preferable with the interferometers adjacent to each other but not in parallel planes, similar to the Bose et. al. setup \cite{bose2017spin}.

Although this is the first  proposal for testing quantum gravity explicitly through GIE with only cold-atom gases, there have been other proposals for testing quantum gravity with cold atoms that use alternative approaches. This includes \cite{Haine_2021} that was discussed above, and \cite{PRXQuantum.2.010325}, where it was demonstrated how non-Gaussianity and Wigner negativity could be used as a witness of quantum gravity, as an alternative to using entanglement. Since negativity has been shown to be equivalent to contextuality, it has been argued that this witness could be  testing the contextuality induced by quantum gravity as opposed to entanglement \cite{PRXQuantum.2.010325}. It was demonstrated that a single, self-gravitating BEC is an ideal system for observing this contextuality witness. In contrast, the experiment we propose here can in principle be applied to non-condensed, ultracold atoms as well as BECs, and uses GIE as evidence of quantum gravity. We hope that this lays the groundwork for further explorations of using GIE of cold atoms to test quantum gravity.    

\begin{acknowledgments}

 All authors acknowledge the support of grant ID 62420 from the John Templeton Foundation. R.H.\ also acknowledges the support of grant ID 62312  from the John Templeton Foundation, as part of the \href{http://www.qiss.fr}{QISS project}.  R.H. thanks Devang Naik for  insightful and useful  discussions.
 

\end{acknowledgments}

\bibliography{references}

\appendix

\section{Gravitational interaction of spheroids} \label{app:oblate}

The coupling $\lambda'_{\alpha \beta}$ in \eqref{eq:lambdap}, can be written as:
\begin{align}
    \lambda'_{\alpha \beta} = \frac{1}{2} \int d^3 \bm{x} \rho_{\alpha} (\bm{x})  \Phi_{\beta}(\bm{x}),
\end{align}
where $\rho_{\alpha} (\bm{x}) :=m  |\psi_{\alpha}(\bm{x})|^2$ is the mass density of the cold atomic gas in mode $\alpha$, and 
\begin{align}
    \Phi_{\beta}(\bm{x}) := \int d^3 \bm{x}' \frac{\rho_{\beta} (\bm{x}')}{|\bm{x} - \bm{x}'|}
\end{align}
is the gravitational potential of the gas in mode $\beta$.

From \cite{Howl_2019}, the gravitational potential outside an oblate  spheroid  in cylindrical coordinates is:
\begin{align}
    \Phi_o = \frac{3 G m}{4 l^3} \Big[ (2 l^2 - D) \sin^{-1} \left(\frac{\sqrt{2} l}{B}\right) + \frac{\sqrt{2} l (A D - l^2 C)}{E B^2} \Big],
\end{align}
where
\begin{align}
    A &:= r^2 + z^2 + \sqrt{z^4 + 2 z^2 (r^2 + l^2) + (l^2 - r^2)^2}\\
    B &:= \sqrt{A+l^2}\\
    C&:= r^2 + 2 z^2\\
    D &:= r^2 - 2 z^2,\\
    E &:= \sqrt{A - l^2},
\end{align}
with $l = \sqrt{a^2 - c^2}$ the focal distance, and $a$ and $c$ respectively the equatorial and polar radii. 

With the gases of the two interferometers two oblate spheriods a distance $d$ apart, and assuming no overlap of the wavefunctions, $\lambda'_{\alpha \beta \neq \alpha}$ can be calculated as \cite{Howl_2019}
\begin{align}
    &\int d^3 \bm{x} \rho_{\alpha} (\bm{x}) \Phi_{\beta} (\bm{x}) = \rho_0 \\ &\times \int_0^{2 \pi} \int^{d+c}_{d-c} \int_0^{a \sqrt{1 - (z - d)^2/c^2}} \,dr \,dz\, d \vartheta\, r\,\Phi_o, 
\end{align}
where $\rho_0 := m / ((4/3) \pi a^2 c)$ is the density of a uniform spheroid. If the two gases where spherical with uniform density, then $\lambda'_{\alpha \beta}$ is simply $ G m^2 / (2d)$. We find  that, assuming an oblate spheroid with same volume as a sphere, this can be improved upon by a factor of $ \approx 3/2$ for  oblate spheroids, assuming $d \approx 2 c$. The ellipticity $e:= l / a$ that achieves this is $e \approx  \sqrt{26/3}/3 \approx 0.98$.

The self-interaction couplings $\lambda'_{\alpha \alpha}$ can be computed in a similar way, resulting in \cite{Howl_2019}:
\begin{align}
    \lambda'_{\alpha \alpha} = \frac{3 G m^2}{5 l} \sin^{-1} e,
\end{align}
which for $e \approx 0.98$, gives $\lambda'_{\alpha \alpha}/\lambda'_{\alpha \beta \neq \alpha} \approx 0.16$. 

Here, we have assumed that the density of the atomic gas is uniform. In practice, the density will depend on the harmonic potential and the strength of the electromagnetic interactions: ranging from a Gaussian profile to a parabolic-like profile \cite{PitaevskiiBook,Howl_2019}. 

\section{Closing both interferometers} \label{sec:closed}

With both interferometers closed, the covariance \eqref{eq:S} is equivalent to $\langle \psi_t |\hat{J}_{\varphi}^{ab} \hat{J}^{cd}_{\varphi'}|\psi_t\rangle - \langle \psi_t| \hat{J}_{\varphi}^{ab}|\psi_t \rangle \langle \psi_t| \hat{J}^{cd}_{\varphi'} |\psi_t \rangle$ before the final beam splitters, where $\varphi$ and $\varphi'$ are the phases induced by two final beam splitters.  This is found to be:
\begin{align}\nonumber
    S &= \frac{1}{8} N^2 \Big[\cos(\mu + \nu) \cos^{2(N-1)}(\lambda + \lambda_s) \\\nonumber &+ \cos(\mu - \nu) \cos^{2(N-1)}(\lambda - \lambda_s)
   \\& - 2 \cos \mu \cos \nu \cos^{2(N-1)}(\lambda_s) \cos^{2 N}(\lambda)\Big].
\end{align}
where $\mu := \varphi - \phi$, $\nu := \varphi' - \phi'$ and we have ignored the electromagnetic interactions between the atoms - to introduce these we just switch $\lambda_s$ with $\gamma$ as defined in the main text.

The variance of this signal is:
\begin{align}\nonumber
 \mathrm{Var}(S) &\approx \frac{1}{\mathbb{M}}\Big[ \langle \left(\hat{J}_{\varphi}^{ab} \hat{J}_{\varphi'}^{cd} \right)^2 \rangle - 2  \langle \left(\hat{J}_{\varphi}^{ab}  \right)^2 \hat{J}_{\varphi'}^{cd} \rangle \langle \hat{J}_{\varphi'}^{cd} \rangle \\ \nonumber&- 2 \langle \left(\hat{J}_{\varphi'}^{cd}  \right)^2 \hat{J}_{\varphi}^{ab} \rangle \langle \hat{J}_{\varphi}^{ab} \rangle  +  \langle \left(\hat{J}_{\varphi}^{ab}  \right)^2\rangle \langle \hat{J}_{\varphi'}^{cd}  \rangle^2 \\ \nonumber&+  \langle \left(\hat{J}_{\varphi'}^{cd}  \right)^2\rangle \langle \hat{J}_{\varphi}^{ab}  \rangle^2 + 6 \langle \hat{J}_{\varphi}^{ab}  \hat{J}_{\varphi'}^{cd}\rangle
 \langle \hat{J}_{\varphi}^{ab}  \rangle
 \langle \hat{J}_{\varphi'}^{cd}  \rangle \\&- 4 \langle \hat{J}_{\varphi}^{ab}  \rangle^2 \langle \hat{J}_{\varphi'}^{cd}  \rangle^2\Big],
\end{align}
where 
\begin{align}\nonumber
     \langle &\left(\hat{J}_{\varphi}^{ab}  \right)^2  \rangle = \frac{1}{8} N \Big(1 + N \\&+ (N-1) \cos(2 (\varphi - \phi)) \cos^{N-2}(2 \lambda_s) \cos^N(2 \lambda)\Big)
     \end{align}
     \begin{align}\nonumber
     \langle &\left(\hat{J}_{\varphi'}^{cd}  \right)^2  \rangle = \frac{1}{8} N \Big(1 + N \\&+ (N-1) \cos(2 (\varphi' - \phi')) \cos^{N-2}(2 \lambda_s) \cos^N(2 \lambda)\Big)
\end{align}
\begin{align}\nonumber
    \langle &\left(\hat{J}_{\varphi}^{ab}  \right)^2 \hat{J}_{\varphi'}^{cd} \rangle = \\\nonumber&\frac{1}{32} N^2 \Bigg[2\cos(\varphi' - \phi' ) \Big(\cos^{N-1} \lambda_s \cos^{N-2} \lambda \, [N + \cos(2 \lambda)]\Big) \\\nonumber&+ (N-1) \Big(
\cos(2 \varphi + \varphi' - 2 \phi - \phi') \cos^{N-2}(2 \lambda_s + \lambda)\\\nonumber&\times \cos^{N-1}(\lambda_s  + 2 \lambda) +\cos(2 \varphi - \varphi' - 2 \phi + \phi') \\&\times\cos^{N-2}(2 \lambda_s - \lambda) \cos^{N-1}(\lambda_s  - 2 \lambda)  \Big) \Bigg]
\end{align}
\begin{align}\nonumber
     \langle &\left(\hat{J}_{\varphi}^{ab}  \right)^2  \left(\hat{J}_{\varphi'}^{cd}  \right)^2 \rangle = \frac{1}{128} N^2 \Big[2(N+1)^2 \\\nonumber&+ (N-1)\Big( 2 \Big(\cos(2(\varphi - \phi)) \\\nonumber&+ \cos(2(\varphi' - \phi'))\Big) \cos^{N-2}(2 \lambda_s) \cos^{N-2}(2 \lambda)[N + \cos(4 \lambda)] \\\nonumber
     &+ (N-1) \cos(2(\varphi+ \varphi'- \phi -\phi')) \cos^{2(N-2)}(2(\lambda_s + \lambda)) \\
     &+ (N-1) \cos(2(\varphi- \varphi'- \phi +\phi')) \cos^{2(N-2)}(2(\lambda_s - \lambda))\Big)\Big].
\end{align}
In the assumption that $\lambda N \ll 1$ and $\lambda_s \approx \lambda$, we find that $\mathrm{SNR} \approx \frac{1}{2} \lambda N$, which is a factor $2$ smaller than the SNR of the scheme considered in the main text.

\section{Entanglement Measure} \label{app:EntMeasure}

Clearly $S$ in \eqref{eq:S} is zero when the state of the atoms is a pure product state. However, classical gravity could be a non-unitary process that causes the two interferometers to couple to each other such that the full state of the system before the final beam splitter becomes mixed:
\begin{align}
\rho_t = \sum_i p_i \rho^{ab}_i \otimes \rho^{cd}_i,
\end{align}
where $0 \leq p_i \leq 1 $ and $\sum_i p_i = 1$. In the case that $\rho^{ab}_i = |\phi_i\rangle_{ab}\,_{ab}{\langle} \phi_i | $ and $\rho^{cd}_i = |\phi'_i\rangle_{cd}\,_{cd}{\langle} \phi'_i | $, with $p_i$ and potentially also $\phi_i$ and $\phi'_i$ dependent on the classical-gravity interaction, $S$ is still zero since:
\begin{align}
    S &= \sum_i p_i \,_{ab}{\langle} \phi_i| \hat{J}^{ab}_y | \phi_i  \rangle_{ab} \,_{cd}{\langle} \phi'_i| \hat{J}^{cd}_z | \phi'_i  \rangle_{cd} \\&\hspace{0.1cm}- \sum_{i,j} p_i p_j \,_{ab}{\langle} \phi_i| \hat{J}^{ab}_y | \phi_i  \rangle_{ab} \,_{cd}{\langle} \phi'_j| \hat{J}^{cd}_z | \phi'_j  \rangle_{cd} \\
    &=0
\end{align}
as $\,_{cd}{\langle} \phi'_i| \hat{J}^{cd}_z | \phi'_i  \rangle_{cd} = 0 $.

More generally, given the symmetry of the setup in Figure \ref{fig:Int}, it would be expected that a classical-gravity interaction will not affect the $50-50$ split of the atoms in each interferometer on average (before the final beam splitter). In this case, even if classical gravity shifted the state of the interferometers $\rho^i_{ab}$ and $\rho^i_{cd}$ away from coherent or squeezed coherent states, we will still obtain $\langle \hat{J}^{cd}_z \rangle = 0$ and thus $S=0$. In the assumption of perfect symmetry, we would also not expect non-gravitational noise to introduce asymmetry in the $50-50$ split. However, even if it did, if it forces the state of the interferometers to be in the number mixed states:
\begin{align}
    \rho^i_{\alpha \beta} = \sum_k c^i_{\alpha \beta \, k} |N-k\rangle_{\alpha} \, _{\alpha}{\langle} N-k | \otimes |k\rangle_{\beta} \, _{\beta}{\langle} k |,
\end{align}
then also $S=0$ for classical gravity since now $\mathrm{Tr} (\rho^i_{\alpha \beta} \hat{J}^{ab}_y) = 0$.

\section{SNR} \label{app:SNR}

From \eqref{eq:Var}, the quantum noise associated with the estimation of the  covariance signal $S$ involves the bi-variate cumulant $\kappa_{22}$, which is \cite{cook1951bi}: 
\begin{align}\nonumber
    \kappa_{22}  &:= \langle \left(\hat{J}_{\varphi}^{ab} \hat{J}_{z}^{cd} \right)^2 \rangle- 2 \langle \left(\hat{J}_{\varphi}^{ab}\right)^2 \hat{J}_{z}^{cd}  \rangle \langle \hat{J}_{z}^{cd}  \rangle
\\\nonumber&-2\langle\hat{J}_{\varphi}^{ab} \left(\hat{J}_{z}^{cd}\right)^2  \rangle \langle \hat{J}_{\varphi}^{ab}  \rangle - \langle \left(\hat{J}_{\varphi}^{ab}\right)^2 \rangle \langle \left(\hat{J}_{z}^{cd}\right)^2 \rangle \\\nonumber&+ 2 \langle \left(\hat{J}_{\varphi}^{ab}\right)^2 \rangle  \langle \hat{J}_{z}^{cd} \rangle^2 + 2 \langle \hat{J}_{\varphi}^{ab} \rangle^2 \langle \left(\hat{J}_{z}^{cd}\right)^2 \rangle   \\\nonumber&+ 8 \langle \hat{J}_{\varphi}^{ab} \hat{J}_z^{cd}\rangle \langle \hat{J}_{\varphi}^{ab} \rangle \langle \hat{J}_{z}^{cd}\rangle
 - 6\langle \hat{J}_{\varphi}^{ab} \rangle^2 \langle \hat{J}_{z}^{cd}\rangle^2 \\\label{eq:k22}&- 2 \langle \hat{J}_{\varphi}^{ab} \hat{J}_z^{cd}\rangle^2.
\end{align}
The noise $\mathrm{Var}(S)$ is then: 
\begin{align} \nonumber
&\mathrm{Var}(S) = \frac{1}{64} \bigg( 2(N+1) \\ \nonumber&+ (N-1) \cos(2 \nu) \cos^{N-2} 2 \lambda_s \cos^{N-2}2 \lambda [2 \\ \nonumber&+ N(\cos 4 \lambda -1)] 
+4 N \cos^2 \nu  \cos^{2N-2}  \lambda_s \cos^{2N-2} \lambda \\ \label{eq:fullVar}&\times [2N \sin^2 \lambda + \cos^2 \lambda - 2]\bigg),
\end{align}
where $\nu := \varphi - \phi $. The SNR is then found through:
\begin{align} \label{eq:fullSNR}
    \mathrm{SNR} = |S|/\sqrt{\mathrm{Var}(S)}, 
\end{align}
with $S$ given by \eqref{eq:fullS}. 

\section{Interrogation time} \label{app:3Body}

Here we estimate the interrogation time of the  atom interferometers by considering the atoms to be erbium and analysing the damping time of erbium atoms due to three-body decay. Loss of cold atoms from traps is dominated by three-body interactions between the atoms, and causes the number density $n$ of the gas to decrease at a rate \cite{SmithBook}:
\begin{align}
    \frac{d n(t)}{dt} = - L n^3(t),
\end{align}
where $L$ is the rate coefficient, which for erbium has been found to be $\lesssim 3 \times 10^{-30}\,\mathrm{cm^6\, s^{-1}}$ \cite{PhysRevLett.112.010404}. Therefore, the time it takes for the number of atoms of an erbium gas to decrease by an order of magnitude (which we take to be the timescale over which the SNR estimate is reliable) is $\gtrsim 2 \times 10^{3} \, \mathrm{s}$ for an initial number density $n\approx 10^{14}\,\mathrm{cm^{-3}}$. This estimate is based on a 3D gas, whereas for 2D and 1D gases, which can be approximated by oblate and prolate spheroidal geometries, three-body losses can be lower \cite{PhysRevA.91.062710,PhysRevA.83.052703,PhysRevA.76.022711}. The estimate of the interrogation time also neglects that the gas is in a superposition of locations, which would be expected to lower the interrogation time. 

\section{Using Feshbach resonances} \label{app:Scattering}

For the example parameter values  in the main text where no squeezing is used, requiring of order $10^{16}$ atoms, the gravitational interaction coupling $2\lambda'$ is of order $10^{-58}\,\mathrm{kg}\,\mathrm{m^2}\,\mathrm{s^{-2}}$. In contrast, ignoring the much weaker magnetic dipole-dipole interactions (MDDIs), the self-interacting electromagnetic coupling $\kappa'$ from \eqref{eq:kappa} for an oblate spheroid gas is $\kappa' = \frac{1}{2} g / ((4/3) a^2 c) = 2 \pi \hbar^2 a_s / ((4/3) a^2 c\, m)$ where we have used $V_{\alpha \alpha} (\bm{x} - \bm{x}') =  g \delta(\bm{x} - \bm{x}')$ , with $g = 4 \pi \hbar^2 a_s/m$   the s-wave coupling constant, and $a_s$ the s-wave scattering length. Setting $a_s = a_0$ the Bohr radius, $\kappa'$ is of order $10^{-50}\,\mathrm{kg}\,\mathrm{m^2}\,\mathrm{s^{-2}}$. From the main text, the electromagnetic start to significantly affect the noise of the signal $S$ once $\kappa'$ becomes 4 orders of magnitude larger than $\lambda'$. Therefore, if Feshbach resonances are used to remove this noise, $a_s$ must be lowered below $a_0$ by 4 or 5 orders of magnitude.  

As stated in the main text, due to the presence of MDDIs, it would likely be preferable to set the s-wave interactions  so that they cancel with the MDDIs, unless $^{88}\protect \mathrm {Sr}$ is ued which has no MDDIs. However, here the same level of noise of the magnetic field would be required to cancel these effects as with setting $a_s$ to $10^{-4} a_0$ as above.

\section{Dependence of purity on gravitational interaction} \label{app:purity}

The density matrix after the interferometers have interacted with each other is $\rho = |\psi_t \rangle \langle \psi_t |$ where $|\psi_t\rangle$ is given by \eqref{eq:StateGravityInt}. Tracing over one of the interferometers (modes $c$ and $d$), we are then left with:
\begin{align}\nonumber
    \rho_{ab} &= \frac{1}{2^{2N}} \sum_{k l} \sqrt{\left( \begin{array}{c} N \\ k \end{array} \right)\left( \begin{array}{c} N \\ l \end{array} \right)} e^{i (k-l) \Phi} e^{i (k^2 - l^2) \gamma} \\&|N- k \rangle_a \, _{a}{\langle}N-l | \otimes |k \rangle_b \, _{b}{\langle} l |  \, \sum_{k'}\sqrt{\left( \begin{array}{c} N \\ k' \end{array} \right)} e^{2 i \lambda k' (k - l)}\\
    &= \frac{1}{2^{N}} \sum_{k l} \sqrt{\left( \begin{array}{c} N \\ k \end{array} \right)\left( \begin{array}{c} N \\ l \end{array} \right)} e^{i (k-l) \theta} e^{i (k^2 - l^2) \gamma} \\&\cos^N(\lambda (k - l))     |N- k \rangle_a \, _{a}{\langle}N-l | \otimes |k \rangle_b \, _{b}{\langle} l |,  
\end{align}
where we have replaced $\lambda_s$ in \eqref{eq:StateGravityInt} with $\gamma:=\lambda_s+\kappa_s - \kappa_d$ to include the electromagnetic interactions within each interferometer (see main text).

Squaring and tracing again, we find the purity of the $ab$ interferometer $\mu := \mathrm{Tr}(\rho^2_{ab})$ to be:
\begin{align}
    \mu &= \frac{1}{2^{2N}} \sum_{k l} \left( \begin{array}{c} N \\ k \end{array} \right)\left( \begin{array}{c} N \\ l \end{array} \right) \cos^{2N}( \lambda(k - l))\\
    &= \frac{1}{2^{2N}} \sum^{2N}_k \left( \begin{array}{c} 2N \\ k \end{array} \right) \cos^{2N}( \lambda (N-k)). 
\end{align}
Note that this is independent of $\gamma$. Therefore, the purity of the state is not affected by the gravitational or electromagnetic self-interactions of each interferometer, and only the gravitational interactions \emph{between} the interferometers, which cause entanglement.

In the assumption that $\lambda N \ll 1$, this approximates:
\begin{align}
    \mu \approx 1 - \frac{1}{2} \lambda^2 N^2.
\end{align}

\section{Noise of isolated interferometer} \label{app:independentInt}

Here we consider a single atom interferometer that is not interacting with another one, and measuring  $\langle \hat{J}_z \rangle$. The interferometer is as the $ab$ interferometer in Figure \ref{fig:Int} in that the two arms are recombined at the end before  $\langle \hat{J}_z \rangle$ is measured. Taking into account the self-interactions of the interferometer, the SNR for this measurement is:
\begin{align}\nonumber
    &\mathrm{SNR} = \sqrt{2N} \cos \delta \cos^{(N-1)} \gamma / \Big( (1 + N) \\\label{eq:SNRSingle}&- 2 N \cos^2 \delta \cos^{2(N-1)} \gamma + (N-1) \cos 2 \nu \cos^{N-2} 2 \gamma\Big)^{\frac{1}{2}},
\end{align}
where $\delta$ is the difference of the phases of the two beam splitters and  and $\gamma:=\lambda_s+\kappa_s - \kappa_d$. 

Given $\lambda N \ll 1$, the SNR of $S$ is:
\begin{align}\nonumber
    &\mathrm{SNR} = \sqrt{2} N \lambda \sin \nu \cos^{(N-1)} \gamma / \Big( (1 + N) \\\nonumber&- 2 N \cos^2 \nu \cos^{2(N-1)} \gamma + (N-1) \cos 2 \nu \cos^{N-2} 2 \gamma\Big)^{\frac{1}{2}},
\end{align}
which has the same dependence on $\gamma$ as \eqref{eq:SNRSingle}, providing another method to isolate the electromagnetic self-interactions of the interferometers.

\end{document}